\pdfoutput=1 

\documentclass[preprint,12pt,a4paper]{elsarticle}

\usepackage{amssymb}

\usepackage{lineno}
\usepackage[utf8]{inputenc}
\usepackage{listings}
\usepackage{xcolor}
\usepackage{color,soul}
\usepackage{multirow}
\definecolor{codegreen}{rgb}{0,0.6,0}
\definecolor{codegray}{rgb}{0.5,0.5,0.5}
\definecolor{codepurple}{rgb}{0.58,0,0.82}
\definecolor{backcolour}{rgb}{0.95,0.95,0.92}

\lstdefinestyle{mystyle}{
    backgroundcolor=\color{backcolour},   
    commentstyle=\color{codegreen},
    keywordstyle=\color{magenta},
    numberstyle=\tiny\color{codegray},
    stringstyle=\color{codepurple},
    basicstyle=\ttfamily\bfseries\footnotesize,
    breakatwhitespace=false,         
    breaklines=true,                 
    captionpos=b,                    
    keepspaces=true,                 
    numbers=left,                    
    numbersep=5pt,                  
    showspaces=false,                
    showstringspaces=false,
    showtabs=false,                  
    tabsize=2
}

\usepackage{slashbox,booktabs}
\usepackage{float}
\usepackage{multirow}
\restylefloat{table}

\newcommand{\squishlist}{
   \begin{list}{$\bullet$}
    { \setlength{\itemsep}{-.1ex}      \setlength{\parsep}{0ex}
      \setlength{\topsep}{0ex}       \setlength{\partopsep}{0ex}
      \setlength{\leftmargin}{.8em} \setlength{\labelwidth}{1em}
      \setlength{\labelsep}{0.5em} } }
\newcommand{\squishend}{\end{list}}
\newcommand{\code}{\textit }
\newcommand{\pkg}{\textbf }

\journal{SoftwareX}

\begin{document}

\begin{frontmatter}



\title{mFLICA: an R package for inferring leadership of coordination from time series}


\author[label2]{Chainarong~Amornbunchornvej\fnref{label4}}

\address[label2]{National Electronics and Computer Technology Center (NECTEC), Pathum Thani, 12120, Thailand}
 \fntext[label4]{Corresponding authors, email: chainarong.amo@nectec.or.th}

\begin{abstract}
Leadership is a process that leaders influence followers to achieve collective goals. One of special cases of leadership is the coordinated pattern initiation. In this context, leaders are initiators who initiate coordinated patterns that everyone follows. Given a set of individual-multivariate time series of real numbers, the mFLICA package provides a framework for R users to infer coordination events within time series, initiators and followers of these coordination events, as well as dynamics of group merging and splitting. The mFLICA package also has a visualization function to make results of leadership inference more understandable. 
\end{abstract}


\end{frontmatter}


\section*{Current code version}

\begin{table}[H]
\begin{tabular}{|l|p{6.5cm}|p{6.5cm}|}
\hline
\textbf{Nr.} & \textbf{Code metadata description} & \textbf{Please fill in this column} \\
\hline
C1 & Current code version & v 0.1.1 \\
\hline
C2 & Permanent link to code/repository used for this code version &  \begin{tiny}$https://github.com/DarkEyes/mFLICA$\end{tiny} \\
\hline
C3 & Code Ocean compute capsule & \begin{tiny}$https://doi.org/10.24433/CO.4248204.v1$\end{tiny}\\
\hline
C4 & Legal Code License   & 	GPL-3 \\
\hline
C5 & Code versioning system used &  git \\
\hline
C6 & Software code languages, tools, and services used &  R, GitHub, TravisCI \\
\hline
C7 & Compilation requirements, operating environments \& dependencies & 64-bit operating system, R (version $\geq$ 3.5.0), R packages: stats, dtw, and ggplot2  \\
\hline
C8 & If available Link to developer documentation/manual &  \begin{tiny}$ https://github.com/DarkEyes/mFLICA$\end{tiny} \\
\hline
C9 & Support email for questions & chainarong.amo@nectec.or.th\\
\hline
\end{tabular}
\caption{Code metadata}
\end{table}








\section{Motivation and significance}
\label{se:motivation}






Leadership is defined as a process that leaders influence a group to achieve collective goals~\cite{hogg2001social,glowacki2015leadership}. One of leadership definitions is pattern initiation. Leaders are initiators who initiate collective patterns (e.g. movement initiation, trends of stock closing prices) that everyone follows~\cite{Amornbunchornvej:2018:CED:3234931.3201406}. Collective patterns or \textit{coordination events} are emerging events of collective actions that aim to reach collective goals~\cite{malone1994interdisciplinary}. In time series context, coordination events occur when there exists some intervals such that some similar pattern occurs in all time series with possibly different time delay for each time series~\cite{Amornbunchornvej:2018:CED:3234931.3201406}. A \textit{leader} of coordination event is a time series that initiates the pattern before others having this similar pattern with arbitrary time delays. 

\begin{tiny}
\begin{table}[]
\caption{Comparison of leadership inference methods}
\label{tab:compMethod}
\begin{tabular}{|p{2.5cm}|p{2.5cm}|p{5cm}|p{3cm}|}
\hline
Method                 & Original domain                 & Pros                                                           & Cons                                                             \\ \hline
mFLICA                 & Coordination initiation         & Detecting multiple group leaders and members of   coordination & Costly time complexity                                           \\ \hline
FLOCK~\cite{andersson2008reporting}                  & Coordination initiation         & Fast and simple method                                         & Cannot detect  multiple   groups in time series \\ \cline{1-3}
LPD~\cite{kjargaard2013time}                    & Leaders inference from movement & Perform well in a small group of trajectories.                 &                                                                  \\ \cline{1-3}
Granger Causality~\cite{granger1969investigating}      & Causal inference                & Perform well in a small group of time series.                  &                                                                  \\ \cline{1-3}
Influence Maximization~\cite{kempe2003maximizing} & Social network analysis         & Detecting multiple leaders and members in social networks      &                                                                  \\ \hline
\end{tabular}
\end{table}
\end{tiny}

The related concepts of leadership inference on time series are Granger causality~\cite{granger1969investigating} (e.g. \pkg{Imtest} package~\cite{GrangerR}) and Transfer Entropy~\cite{schreiber-prl00,BEHRENDT2019100265} (e.g. \pkg{RTransferEntropy} package~\cite{BEHRENDT2019100265}).  Both techniques can be used to infer whether time series $A$ is a predictor of time series $B$, which is similar to the following relation concept in leadership inference.  Nevertheless, leadership inference aims to identity patterns {(e.g. moving to the same trajectory)} that distributes among time series and their initiators (leaders) rather than finding predictors. {There are many leadership methods in the literature (see Table}~\ref{tab:compMethod}{). However, there is no leadership-inference R software package in the literature that can detecting multiple group leaders and members of coordination yet. For more details regarding the literature review of leadership methodologies, please see}~\cite{Amornbunchornvej:2018:CED:3234931.3201406,mFLICASDM18}.  

To fill the gap in the literature, in this paper, I developed an R package for leadership inference in R~\cite{Rprog} on The Comprehensive R Archive Network (CRAN)~\cite{RmFLICA}: \pkg{mFLICA}. {The methodology of this framework is based on}~\cite{Amornbunchornvej:2018:CED:3234931.3201406,mFLICASDM18}{, which has been peer-reviewed and tested on both noisy simulation datasets and real-world datasets.} \pkg{mFLICA} is a framework that is capable of:

\squishlist
\item {\bf Inferring coordination events:} the framework can infer and visualize coordination intervals that have high degrees of coordination; and
\item {\bf Inferring dynamics of leaders and followers:} the framework can infer leaders of coordination and their followers that can be changed over time. 
\squishend
The \pkg{mFLICA} package provides scientists opportunities to analyze and generate scientific hypotheses on coordinated activities that can be tested statistically and in the field. {Note that for the details of algorithms in functions and definitions related to this work, see the supplementary document and}~\cite{mFLICASDM18,Amornbunchornvej:2018:CED:3234931.3201406} {for more details regarding the performance of methodology and its theoretical properties. }

\subsection{Limitation}
{The} \pkg{mFLICA} {package has been built based on Dynamic Time Warping (DTW)}~\cite{sakoe1978dynamic}. {Hence, it is an optimization framework that can be used to detect leadership patterns. However, even though the package cannot provide any statistics as outputs, instead, users can derive statistics from the framework outputs. For instance, a confidence interval of a specific individual being leaders can be derived by bootstrapping time series of leaders. }

{Another assumption is that the framework assumes that no external influences that cannot be found in the datasets. This implies that if the datasets contains partial information without including external influence, leaders found by the framework might not be the true leaders. Hence, users should carefully interpret their results with this assumption. }



\section{Software description}
\label{sec:softdes}
I provide details of mFLICA system architecture in Section~\ref{sec:SWArchitecture}, then I describe software functionality in Section~\ref{sec:SWfunc}. 
\begin{figure}[t!]
\centering
\includegraphics[width=1\columnwidth]{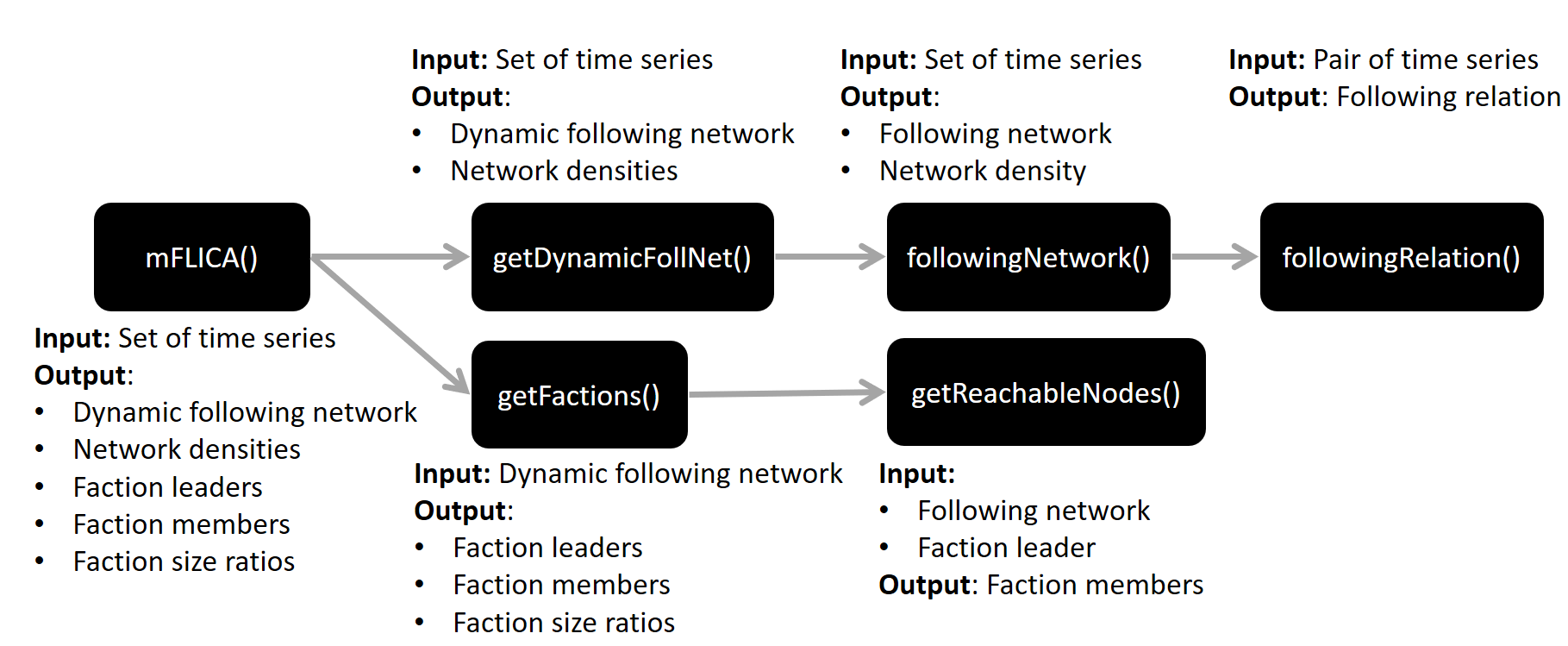}
\caption{Software architecture of mFLICA. Given a set of time series, mFLICA infer following networks, faction leaders and members, degrees of coordination over time, as well as related visualization.}
\label{fig:mainArchitecture}
\end{figure}

\subsection{Software Architecture}
\label{sec:SWArchitecture}
Given a set of time series and related parameters as inputs, \textbf{mFLICA} infers following networks, faction leaders and members, degrees of coordination over time, as well as related visualization.

Figure~\ref{fig:mainArchitecture} provides an overview of the package architecture. The main function is {mFLICA()} that calls two functions: {getDynamicFollNet()} and {getFactions()}. The {getDynamicFollNet()} is used to infer a dynamic following network from a set of time series, while {getFactions()} is used to infer faction leaders and faction members for each time step in a dynamic following network. In {getDynamicFollNet()}, it calls {followingNetwork()} for inferring a following network for each time intervals to create a dynamic following network. The {followingNetwork()} function uses {followingRelation()} as a main engine to infer a following relation between a pair of time series to build a following network. Lastly, {getFactions()} calls {getReachableNodes()} to find faction members, which are nodes that have directed path(s) to the faction leader.

\subsection{Software Functionalities with Examples}
\label{sec:SWfunc}

The main tasks of mFLICA are 1) inferring a dynamic following network, and 2) inferring faction leaders and members as well as leadership dynamics. 

In this paper, I use a simulated dataset $TS$ that contains 30 time series of movement from~\cite{mFLICASDM18} to demonstrate in examples of using \pkg{mFLICA} in leadership inference tasks. The dataset consists of two-dimensional time series of 30 individuals moving along the x-axis. The time series length is 800 time steps. There are three coordination events during the time interval [1,200] leading by individual ID1, the time interval [201,400] leading by ID2, and the time interval [400,600] leading by ID3. The dataset is included in this package.

\subsubsection{Inferring following relations} 
\begin{figure}[ht!]
\centering
\includegraphics[width=1\columnwidth]{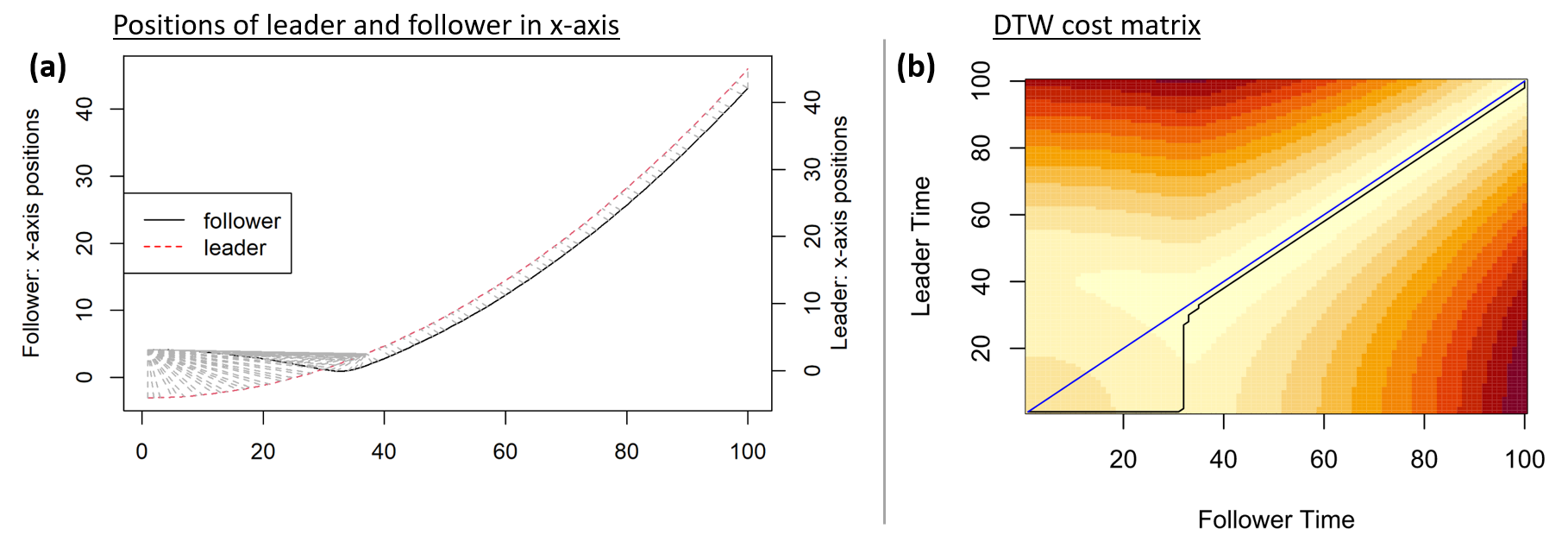}
\caption{(a) Leader and follower time series in x-axis. These time series of positions are generated based on the movement of individuals in a two-dimensional plane where a leader moved along x-axis.  A follower moved toward its leader in this plane. Both leader and follower have almost the same values in y-axis. (b) DTW cost matrix where darker-color shades represent higher distance. The black line is the optimal warping path between leader and follower, while the blue line is the diagonal line.}
\label{fig:FollTSEx}
\end{figure}

To infer a following relation between two time series, I deploy dynamic time warping (DTW) package~\cite{Rdtw} to analyze an optimal warping path between two time series. Figure~\ref{fig:FollTSEx} shows simulated time series of movement from~\cite{mFLICASDM18}. In this event, a leader was moving toward x-axis while the follower followed its leader after some time delay.  A degree of following relation can be defined below.

\begin{equation}
\label{eq:foll}
    s(P_{L,F}) = \frac{\sum_{i \in P_{L,F}} (\textit{sign}(c_i-r_i))}{|P_{L,F}|}
\end{equation}
 
 Where $P_{L,F}$ is the optimal warping path of leader $L$, and follower $F$ inferred by DTW and $s(P_{L,F}) \in [-1,1]$. Given a threshold $\sigma \in [0,1]$, if $s(P_{L,F}) \in [-1,-\sigma]$, then $L$ follows $F$.  If $s(P_{L,F}) \in [\sigma,1]$, then $F$ follows $L$. Otherwise, there is no following relation for $s(P_{L,F}) \in (-\sigma,\sigma)$. 
 
 In the next example, I also deploy two time series in $TS$. In this dataset, $TS[1,1:100,]$  is a time series of leader while $TS[2,1:100,]$ is a time series of follower. I use only the first 100 steps of time interval in this example. I run the code below for computing the optimal warping path between leader and follower.

\begin{lstlisting}[language=R]
R>library(mFLICA)
R>leader<-mFLICA::TS[1,1:100,] # Optimal warping path: obj$index2
R>follower<-mFLICA::TS[2,1:100,] # Optimal warping path: obj$index1
R>obj<-dtw(x=follower,y=leader,k=TRUE)   # run dtw from 'dtw' package
\end{lstlisting}

I called $dtw()$ function and recorded the result in $obj$ where  $obj\$index1$ contains optimal warping path of a follower, and  $obj\$index2$ contains optimal warping path of a leader. This means the follower at time $obj\$index1[i]$ is matched (most similar w.r.t. DTW matching) with the leader at time $obj\$index2[i]$. Then, I compute the average number of time steps until the follower reached the leader's previous positions.

\begin{lstlisting}[language=R]
R> mean( obj$index1 - obj$index2)
[1] 8.238462
\end{lstlisting}

On average, the follower required eight time steps to reach its leader. Next, I calculate $s(P_{L,F})$ in Eq.~\ref{eq:foll}.

\begin{lstlisting}[language=R]
R>mean(sign( obj$index1 - obj$index2) ) 
[1] 0.9846154
\end{lstlisting}

This implies that there is a high degree of following relation between leader and follower ($s(P_{L,F})\approx 0.98$). The $followingRelation()$ function in the package is for computing $s(P_{L,F})$.  
I deploy Sakoe-Chiba Banding~\cite{sakoe1978dynamic} for speeding up DTW computation. The limitation of band  can be set via $lagWindow$ parameter. In this case, I set the band parameter at 10\% of the time series length ($lagWindow=0.1$). 

\begin{lstlisting}[language=R]
R> mFLICA::followingRelation(Y=follower,X=leader,lagWindow=0.1)$follVal
[1] 0.99
\end{lstlisting}

I have $s(P_{L,F}) = 0.99$ in this example. 

\subsubsection{Inferring following network}  
The $followingNetwork()$ function is used to infer an adjacency matrix of a following network. The code below is used to infer adjacency matrices by using a set of simulated time series $TS$, which contains 30 time series. The low-coordination interval $[1,60]$ and high-coordination interval $[61,120]$ are chosen in the example.  I set $\sigma = 0,5$ for this example.

\begin{lstlisting}[language=R]
R>library(mFLICA)
R>mat1<-followingNetwork(TS=TS[,1:60,],sigma=0.5)$adjWeightedMat
R>mat2<-followingNetwork(TS=TS[,61:120,],sigma=0.5)$adjWeightedMat
\end{lstlisting}

\begin{figure}[ht!]
\centering
\includegraphics[width=1\columnwidth]{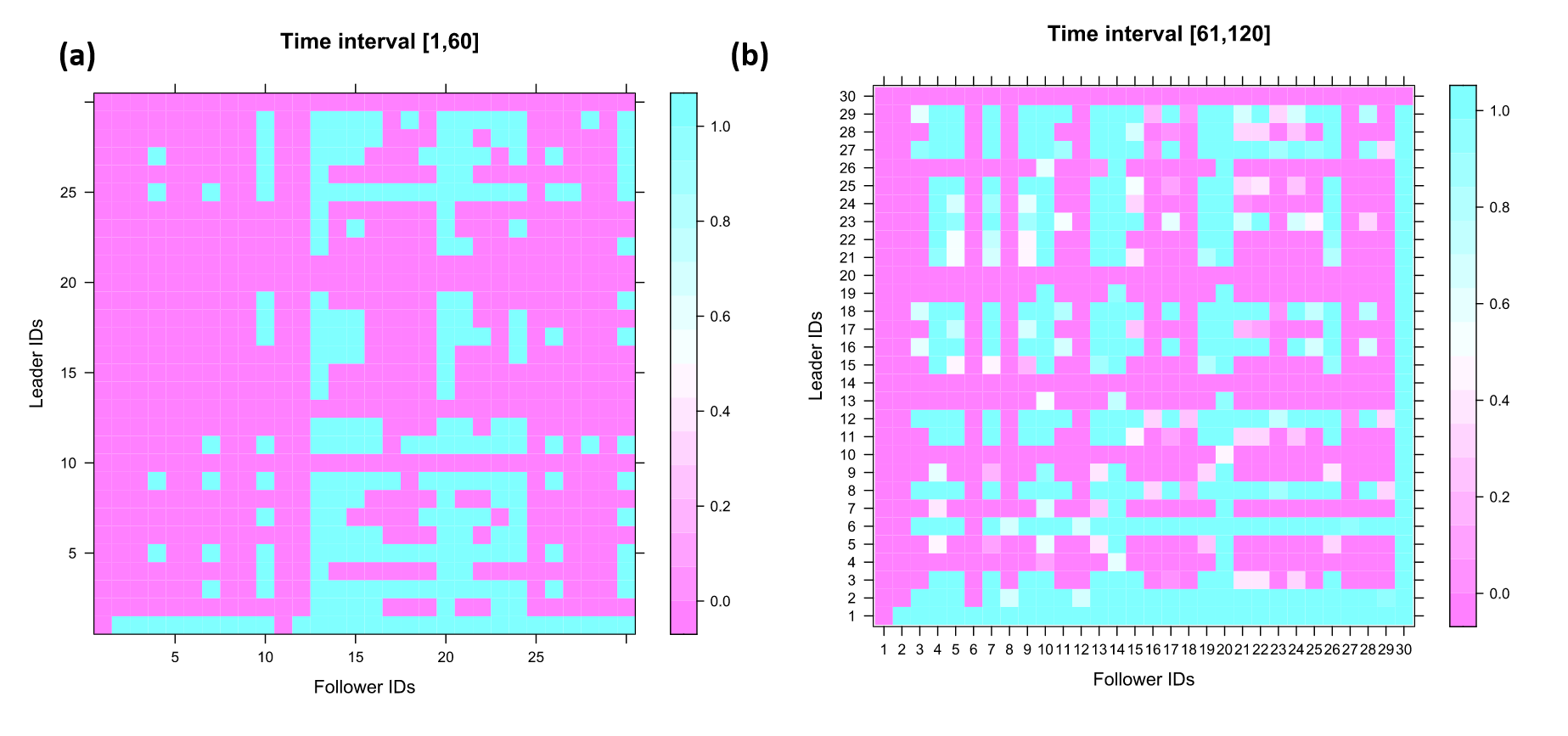}
\caption{Weighted adjacency matrices of the following networks from difference time intervals. Elements in matrices represent degrees of following derived from $s(P_{L,F})$ in Eq.~\ref{eq:foll} where leaders are rows and followers are columns (e.g. $(i,j)=0.5$ implies $j$ follows $i$ with degree $0.5$). A higher value (light blue) implies a higher degree of following relation. (a) The matrix from the interval [1,60], which has low degrees of coordination. (b) The matrix from the interval [61,120], which has high degrees of coordination leading by ID1. }
\label{fig:FollNetEx}
\end{figure}

Figure~\ref{fig:FollNetEx} illustrates the adjacency matrices from both intervals. The weighted adjacency matrix {mat1} at Figure~\ref{fig:FollNetEx} (a) is computed from the time interval $[1,60]$ when the group initiated movement. In Figure~\ref{fig:FollNetEx} (b),  the weighted adjacency matrix {mat2} is computed from  the time interval $[61,120]$ when everyone followed its leader ID1, which implies it is a high-coordination event.  {mFLICA} provides {getADJNetDen()} for computing a network density from an adjacency matrix. Based on the result, {mat1} has a lower network density than {mat2}'s network density. The network densities can be computed below.  

\begin{lstlisting}[language=R]
R>getADJNetDen(mat1)
[1] 0.5559004
R>getADJNetDen(mat2)
[1] 0.7961686
\end{lstlisting}

In Figure~\ref{fig:FollNetEx} (b), in the row of ID1, all individuals have high degrees of following ID1, which implies that ID1 is a leader in this interval. In contrast, there are no individuals followed by the majority in Figure~\ref{fig:FollNetEx} (a), which implies that this interval has low degrees of coordination.

\subsubsection{Inferring dynamic following network} 
In this part, I use the set of simulated time series \code{TS}, which has the time length at 800 time steps. In this dataset, there are three coordination events: [1,200], [201,400], and [401,600]. I set the time window $\omega=60$, the time shift $\delta=6$, and the threshold $\sigma=0.5$. The next commands are used to infer a dynamic following network of \code{TS}.

\begin{lstlisting}[language=R]
R>library(mFLICA)
R>obj1<-getDynamicFollNet(TS=mFLICA::TS[,1:800,],timeWindow=60,timeShift = 6,sigma=0.5)
\end{lstlisting}

Suppose I want to know the following degree for ID19 follows ID1 at time step 150, I can use the command below.
\begin{lstlisting}[language=R]
R>obj1$dyNetWeightedMat[19,1,150]
[1] 0.9833333 # the following degree for ID19 follows ID1 at time step 150
\end{lstlisting}
I can query the network density at time step 150 with the command below.
\begin{lstlisting}[language=R]
R>obj1$dyNetWeightedDensityVec[150]
[1] 0.7755939
\end{lstlisting}

The time series of network densities can be plotted using the \code{plotMultipleTimeSeries} function below. 

\begin{lstlisting}[language=R]
R>plotMultipleTimeSeries(TS=obj1$dyNetWeightedDensityVec, strTitle="Network Dnesity")
\end{lstlisting}

\begin{figure}[ht!]
\centering
\includegraphics[width=1\columnwidth]{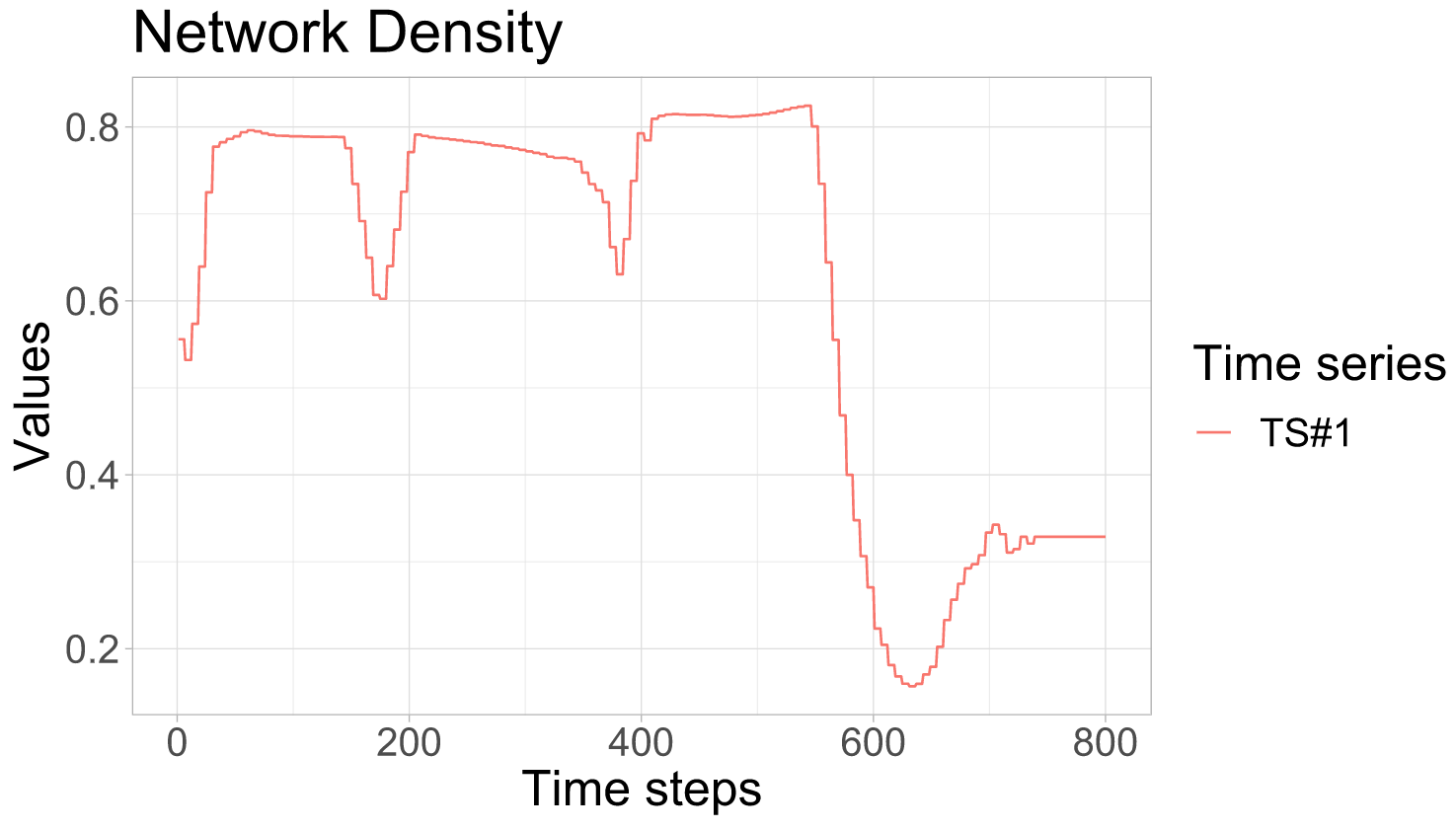}
\caption{Time series of network densities of a dynamic following network from simulated time series \code{TS} in \pkg{mFLICA} package. The plot shows that there are three coordination events that have high network densities (high degrees of coordination): [1,200], [201,400], and [401,600], which are consistent with our ground truth. }
\label{fig:DyNetDen}
\end{figure}

Figure~\ref{fig:DyNetDen} shows the result of the plot. The plot shows that there are three coordination events that have high network densities (high degrees of coordination): [1,200], [201,400], and [401,600], which are consistent with our ground truth.

\subsubsection{Inferring leadership dynamics}
I use the interval [25,45] in the simulation dataset to demonstrate the time when there are more than one factions occur simultaneously. After having a following network, \code{getFactions()} takes a binary version of adjacency matrix as its input.
\begin{lstlisting}[language=R]
R>library(mFLICA)
R>mat1 <-followingNetwork(TS=TS[,25:45,], sigma =0.95)$adjBinMat
R>out<-getFactions(adjMat=mat1)
R>out$leaders # show leader IDs
[1] 1 11
\end{lstlisting}
The code above shows that there are two faction leaders in the interval [25,45]: ID1 and ID11. This implies that there are two factions. The next step is to query faction members of ID1's faction as well as its faction size ratio.
\begin{lstlisting}[language=R]
R>L1<-out$leaders[1] # leader ID1
R>out$factionMembers[[1]] # show faction members
[1]  1  2  3  4  5  6  7  8  9 10 12 13 14 15 16 17 18 19 20 21 22 23 24 25 26 27 28 29 30 
R>out$factionSizeRatio[L1] # show faction size ratio
[1] 0.5034483
\end{lstlisting}
Note that a leader is also a faction member itself. Since there are 30 individuals, almost everyone is a member of ID1's faction. However, the faction size ratio at 0.5 indicates that faction members are not coordinated following the same pattern yet. The next one is the code for querying details about a faction leading by ID11.

\begin{lstlisting}[language=R]
R>L1<-out$leaders[2] # leader ID11
R>out$factionMembers[[2]] # show faction members
[1]  11  7 10 13 14 15 16 18 19 20 21 22 23 24 26 28 30 
R>out$factionSizeRatio[L2] # show faction size ratio
[1] 0.1632184
\end{lstlisting}
We can see that there are a fewer number of members in this faction. Note that one individual can belong to more than one faction since the individual might follow some pattern that seems partially similar to several leaders' patterns.

Next, I show how to use \pkg{mFLICA} to infer dynamics of factions. In other words, I would like to find changes of faction members and faction leaders over time. Given a set of time series \code{TS} as an input along with related parameters: time window $\omega=60$, time shift $\delta=6$, and the threshold $\sigma=0.5$, we run \code{mFLICA()} below.
\begin{lstlisting}[language=R]
R>library(mFLICA)
R>obj1<-mFLICA(TS=mFLICA::TS[,1:800,],timeWindow=60,timeShift = 6,sigma=0.5)
\end{lstlisting}

All results of faction inference are in \code{obj1}. Here, we focus on a set of time series of faction size ratios \code{obj1\$factionSizeRatioTimeSeries} where \code{obj1\$factionSizeRatioTimeSeries[i,t]} is a faction size ratio of a faction leading by ID\code{i} at time \code{t}. We can plot the time series of faction size ratios using \code{plotMultipleTimeSeries} below.

\begin{lstlisting}[language=R]
R>plotMultipleTimeSeries(TS=obj1$factionSizeRatioTimeSeries, strTitle="Faction Size Ratios")
\end{lstlisting}

\begin{figure}[ht!]
\centering
\includegraphics[width=1\columnwidth]{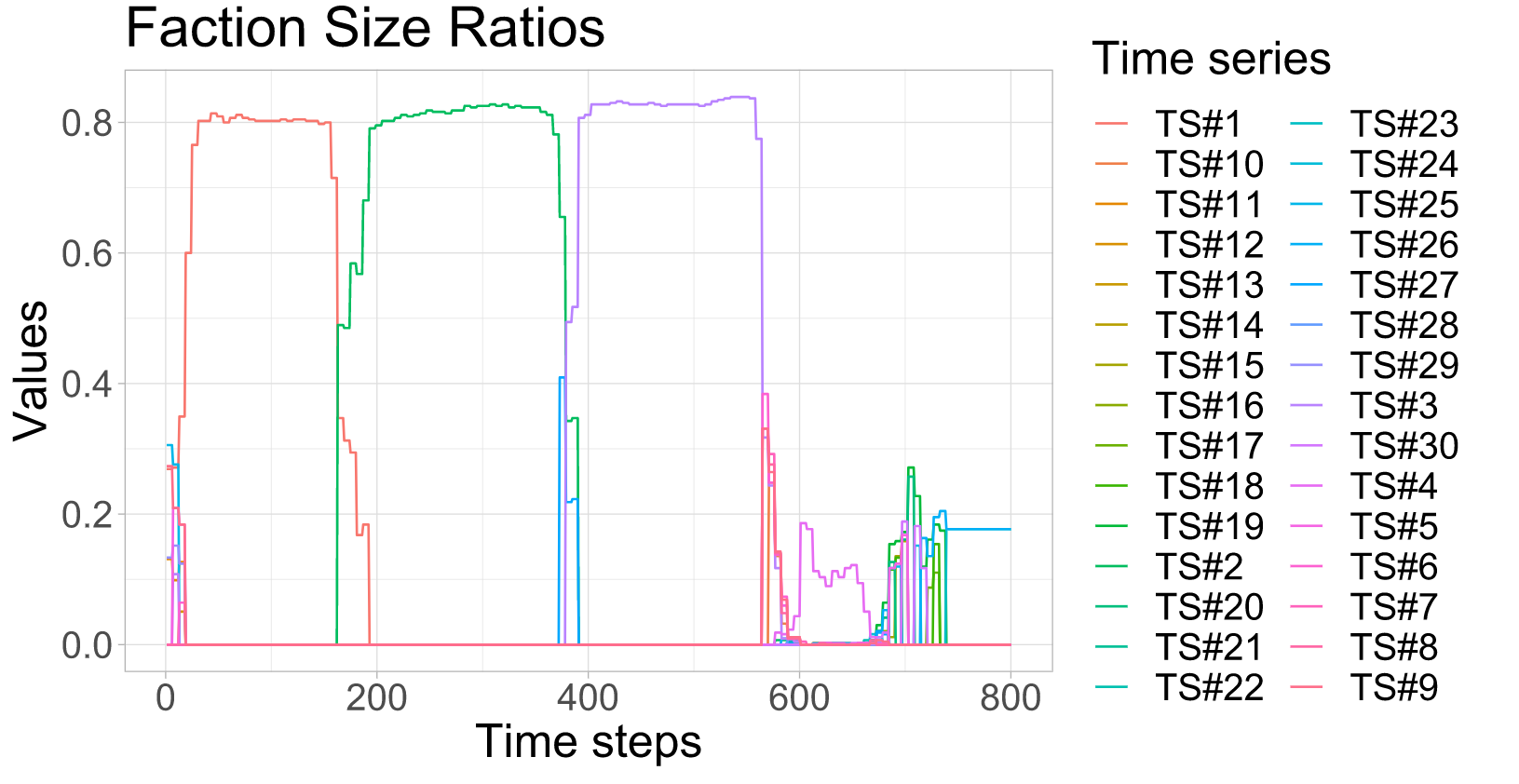}
\caption{Time series of faction size ratios. Each time series of faction sizes ratios is represented by ID (TS\#ID) of its faction leader.  }
\label{fig:FactionDy}
\end{figure}

The result of the plot is in Figure~\ref{fig:FactionDy}. According to the ground truth, there are three coordination events. First, during the time interval [1,200], ID1 is a sole leader who leads its faction of 30 individuals. Then, ID2 leads the faction for the time interval [201,400]. Afterward, ID3 leads the faction during the interval [401,600]. Finally, the group slows down and stop moving. The result in Figure~\ref{fig:FactionDy} reflexes this ground truth. ID1 has its high faction size ratios during [1,200], ID2's faction continues to have high faction size ratios during [201,400]. Lastly, ID3's faction has high faction size ratios during [401,600]. No factions have high faction size ratios during [601,800].
\section{Impact}
\label{sec:impact}


In the past, many social science questions including leadership were unable to be answers with quantitative approaches due to the lack of resource and data. 
Currently, because of innovation and technology developments, data from both online social network and real-world sensors of behaviors of human, animal, or even man-made systems are available. These datasets open opportunities for researchers to ask questions and gain insight about collective behaviors quantitatively. In leadership inference, mFLICA package enables computer scientists, social scientists, and researchers to quantitatively test hypotheses regarding leadership of coordination.

In the commercial realm, mFLICA is able to support companies to measure influence of their products among customers from customers' records. Understanding effects of products is a crucial part for companies to gain or loss profits. 

In the research realm, there are currently some examples of  new research regarding the potential of utilizing mFLICA in the literature. In online social behavior analysis, there is a recent work in~\cite{santagiustina2021unfolding} that used mFLICA to analyze time series of records from Twitter's online users for  obtaining the structure of arguments in online debates about  "no-deal" Brexit. In the animal behaviors, the work in~\cite{amornbunchornvej2020framework} utilized mFLICA to gain the results from GPS trajectories of 26 baboons and found that baboons do not follow any particular dictator but they follow the group, which is consistent with the biological result in~\cite{strandburg2015shared}. Additionally, in social science, the work in~\cite{COOK2020101296} stated that a leadership-inference framework (e.g. mFLICA) has a potential to be used for obtaining causal relationships and influence between people.

\section{Conclusions}
\label{sec:conclusion}

In this paper, the details of \pkg{mFLICA} package for inferring leadership of coordination from time series are provides. Leaders are defined as individuals who initiate some patterns and others follow the same patterns with some time delays. A following relation between time series can be detected by analyzing an optimal warping path of Dynamic Time Warping (DTW), which is the main component that \pkg{mFLICA} deploys. 

Given a set of time series and related parameters, the \pkg{mFLICA} package can infer a following relation between two time series, following networks, faction leaders, faction members, degrees of coordination, and faction size ratios for each time step. 

The network densities inferred by \pkg{mFLICA} tell us regarding the magnitude of coordination: how many time-series individuals follow the same pattern in a given time interval. The faction size ratios provide information regarding faction dynamics; the changes of faction leaders, and/or faction members over time. We provided the examples of how to use \pkg{mFLICA} for solving many tasks in leadership inference. Our framework can be applied to any multivariate time series.

\section{Conflict of Interest}


I wish to confirm that there are no known conflicts of interest associated with this publication and there has been no significant financial support for this work that could have influenced its outcome.

 \section*{Acknowledgements}
{The author would like to acknowledge the support from NSF III-1514126 in the development of the methodology used in this work.}

\appendix

\section{The definitions and pseudo code for inferring leadership of coordination}
Supplementary material related to this article can be found online at https://doi.org/10.1016/j.softx.2021.100781.
 \bibliographystyle{elsarticle-num} 
 \bibliography{ChaiRef}






\end{document}